\documentclass[prl,aps,amssymb,showpacs,twocolumn]{revtex4-1}
\usepackage{amsmath}
\usepackage{amssymb}
\usepackage{amsthm}
\usepackage{amsfonts}
\usepackage{listings}
\usepackage{enumerate}
\usepackage{latexsym}
\usepackage[dvips]{graphicx}
\usepackage{psfrag}
\usepackage{bm}
\usepackage{graphicx}
\usepackage{subfigure}

\newcommand{\beq}{\begin{equation}}
\newcommand{\eneq}{\end{equation}}

\newcommand{\braket}[2]{\left\langle #1 | #2 \right\rangle}
\newcommand{\bra}[1]{\left\langle#1\right|}
\newcommand{\ket}[1]{\left|#1\right\rangle}

\newcommand{\pref}[1]{(\ref{#1})}

\def\ie{{\it i.e.},\ }

\input{epsf}

\begin{document}

\tolerance 10000

\newcommand{\vk}{{\bf k}}


\title{Extracting Excitations From Model State Entanglement}

\author{A. Sterdyniak$^{1}$, N. Regnault$^{1}$ and  B.A. Bernevig$^{2}$} 
\affiliation{$^1$ Laboratoire Pierre Aigrain, ENS and CNRS, 24 rue Lhomond, 75005 Paris, France}
\affiliation{$^2$ Department of Physics,
  Princeton University, Princeton, NJ 08544} 

\begin{abstract}
We extend the concept of entanglement spectrum from the geometrical to the particle bipartite partition. We apply this to several  Fractional Quantum Hall (FQH)  wavefunctions on both sphere and torus geometries to show that this new type of entanglement spectra completely reveals the physics of bulk quasihole excitations. While this is easily understood when a local Hamiltonian for the model state exists, we show that the quasiholes wavefunctions are encoded within the model state even when such a Hamiltonian is not known. As a nontrivial  example, we look at Jain's composite fermion states and obtain their quasiholes directly from the model state wavefunction. We reach similar conclusions for wavefunctions described by Jack polynomials. 
\end{abstract}

\date{\today}


\maketitle

Topological phases are highly nontrivial states of matter whose complete characterization has, despite intense effort, remained elusive.  The parade example of a topological ordered phase, and the only one so far realized in experiments, is the Fractional Quantum Hall (FQH) effect. Active ongoing efforts to understand the physics of these systems focus on several issues: an important long-standing research direction in topological phases focuses on  finding the best numerical techniques to identify topological order in realistic systems. Due to the absence of a local order parameter, this is a highly nontrivial task.  A related recent research direction has focused on extracting as much information as possible about a topological phase -- including information about its excitations -- purely from its ground state wavefunction. The deep conceptual question is whether the ground state of a generic Hamiltonian encodes the complete information about the universality class of a topologically ordered system, and if so, what is the best way to extract it.

Towards this end, it has been recently proposed and numerically substantiated that the physical properties of the FQH edge can be obtained from the ground state using the entanglement spectrum (ES)\cite{li-08prl010504}. For a single non-degenerated ground state $\ket{\Psi}$, ES can be defined through the Schmidt decomposition of $\ket{\Psi}$ in two regions $A$, $B$ (not necessarily spatial):

\begin{equation}
\ket{\Psi}=\sum_i e^{-\xi_i / 2} \ket{\Psi^A_i} \otimes \ket{\Psi^B_i}
\label{schmidt}
\end{equation}

\noindent where $\braket{\Psi^A_i}{\Psi^A_j}=\braket{\Psi^B_i}{\Psi^B_j}=\delta_{i,j}$. The $\exp(-\xi_i)$ and $ \ket{\Psi^A_i}$ are the eigenvalues and eigenstates of the reduced density matrix $\rho_A={\rm Tr}_B \rho$, where $\rho=\ket{\Psi}\bra{\Psi}$ is the total density matrix. Writing $\rho_A=\exp(-H)$, $\xi_i$ and $\ket{\Psi^A_i}$ can be regarded as the eigenvalues and eigenstates of a fictitious hamiltonian $H$. The ES is the spectrum associated to $H$.

It has been shown numerically on a case by case basis \cite{li-08prl010504,regnault-09prl016801} that the counting of the low energy part of the $\xi_i$'s in the ES matches the counting of the edge modes of the respective FQH state. In the case of a realistic system whose ground state is close to a model FQH wavefunction, such as the Coulomb ground state in the lowest Landau level (LLL) at filling factor $\nu=1/3$ (close to the Laughlin state), an entanglement gap can be defined separating a low energy part matching the model state $\xi$'s and a non-universal high energy spectrum \cite{thomale-10pr180502}. The entanglement gap is numerically conjectured to remain finite in the thermodynamic (TD) limit if the realistic system is in the same universality class as the model wavefunction. The ES has also been applied to other systems such as quantum spin systems \cite{thomale-09es, PhysRevLett.105.077202} or topological insulators \cite{haldane-march2009, fidkowski2010, turner-09cm0909,PhysRevLett.105.115501}. 
 
\begin{figure}[t]
\includegraphics[width=0.7\linewidth]{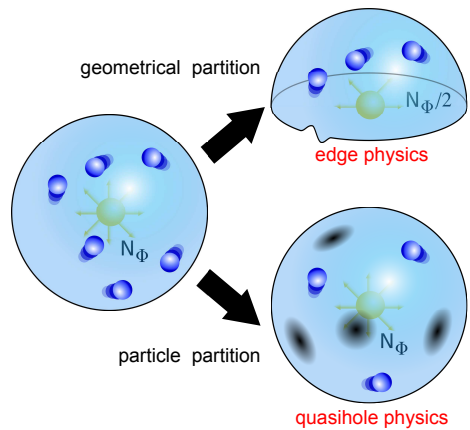}
\caption{Two types of bipartite partition that can be done on the ground state. The geometrical or orbital cut (right top corner) probes the edge physics; the particle cut (right bottom corner) allows access to the bulk qh excitations.}
\label{partition}
\end{figure}

The ES in \cite{li-08prl010504} is related to the previously introduced \cite{haque-07prl60401} FQH geometrical bipartite entanglement entropy (EE) $S = -\sum_i \xi_i \log\xi_i$.  References \cite{haque-07prl60401,haque-07prb125310} also introduced a different type of EE based on a particle rather than geometrical cut.  In their case, the $A$ part is a subset of the total particles, while the $B$ part is the remaining particles. The geometry is left untouched. In a similar way \cite{li-08prl010504} extended the geometrical bipartite EE to the GES (geometrical ES), we propose to define a particle ES (PES) from the particle bipartite EE of \cite{haque-07prl60401} and analyze its behavior for both model and realistic FQH ground states. For the FQH state on the sphere, the eigenstates of the PES will preserve the sphere, its radius and the number of flux quanta, but with a smaller number particles living on it than in the ground state. Intuitively, this situation corresponds to nucleating bulk quasihole (qh) excitations. Fig. \ref{partition} sketches this situation. The PES provides us with both the correct counting of, and the actual, quasihole eigenstates of a FQH state. We show that the PES gives correct results even for states without a known local Hamiltonian, such as the Composite Fermion (CF) or some Jack wavefunctions. We then extend the PES to the torus geometry, with similar results.

In this paper, we mostly focus on FQH states on the sphere with $N$ number of particles under $N_\phi$  flux quanta, and label states by their total angular momentum $L$ and its projection $L_z$. The reduced density matrix in the bipartite particle partition preserves the symmetry of the original state. We decompose any one body operator ${\cal O}$ into ${\cal O}_A+{\cal O}_B$ where ${\cal O}_A$ (${\cal O}_B$) only acts on the $A$ ($B$) group of particles. If $[{\cal O}, \rho]=0$ (true for ${\cal{O}}= L^{\pm, z}$ when the state $\ket{\Psi}$ has $L=0$), we also have $0=\text{Tr}_B [{\cal O}_A, \rho] + \text{Tr}_B [{\cal O}_B, \rho] =[{\cal O}_A, \text{Tr}_B \rho]= [{\cal O}_A, \rho_A] $ as the trace over the $B$ degrees of freedom of a commutator operator in the $B$ part vanishes. We classify the $\rho_A$ eigenvalues by their total angular momentum $L_A$ and its projection $L_{z,A}$ and plot the PES as a function of $L_A$ (instead of $L_{z,A}$ for the GES) to remove the multiplet degeneracy. Numerically, the PES involves the diagonalization of larger matrices than those of the GES.

\begin{figure*}[t]
  \begin{minipage}[l]{0.32\linewidth}
    \includegraphics[width=\linewidth]{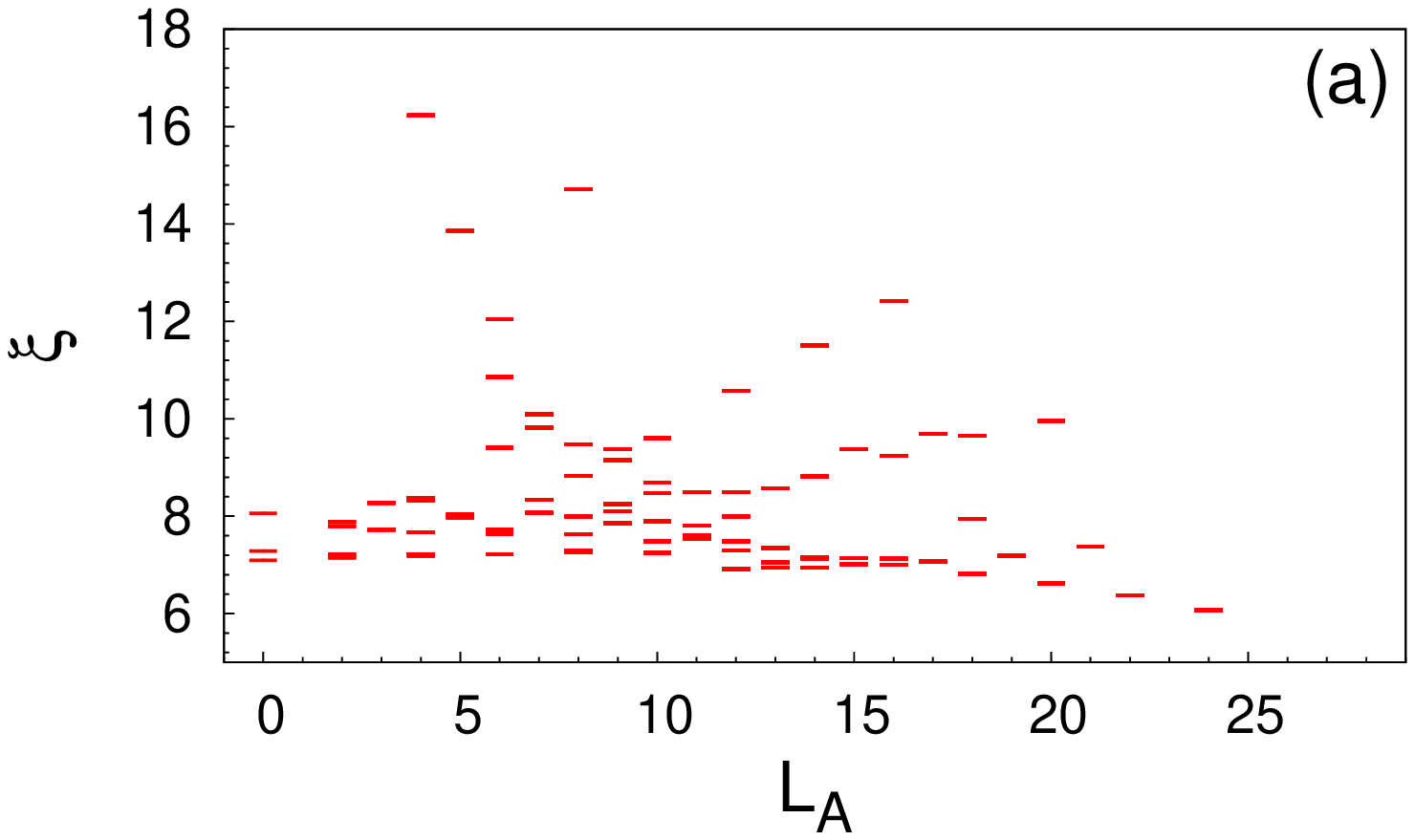}
  \end{minipage}
\hspace{1pt}
  \begin{minipage}[l]{0.32\linewidth}
    \includegraphics[width=\linewidth]{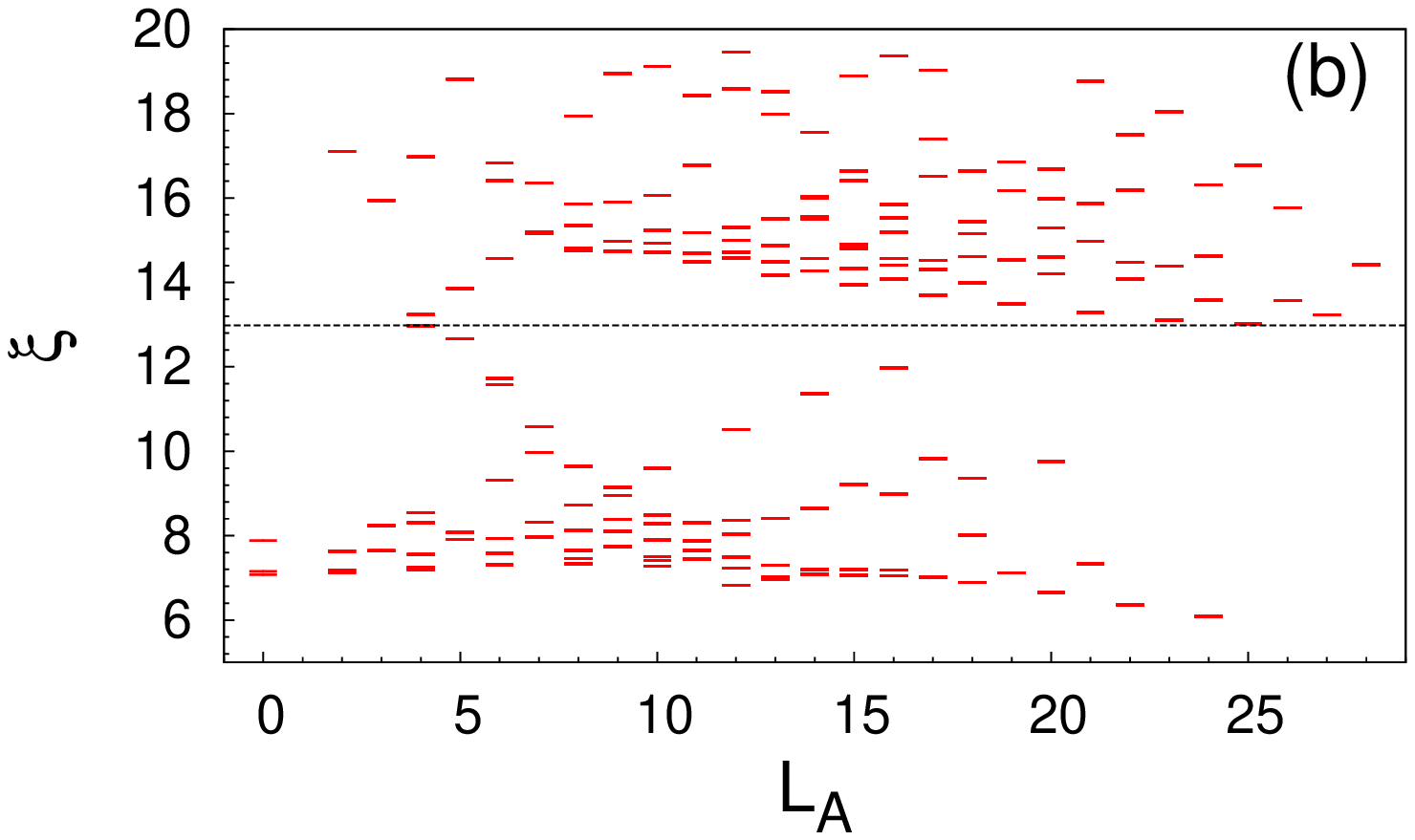}
  \end{minipage}
\hspace{1pt}
  \begin{minipage}[l]{0.32\linewidth}
    \includegraphics[width=\linewidth]{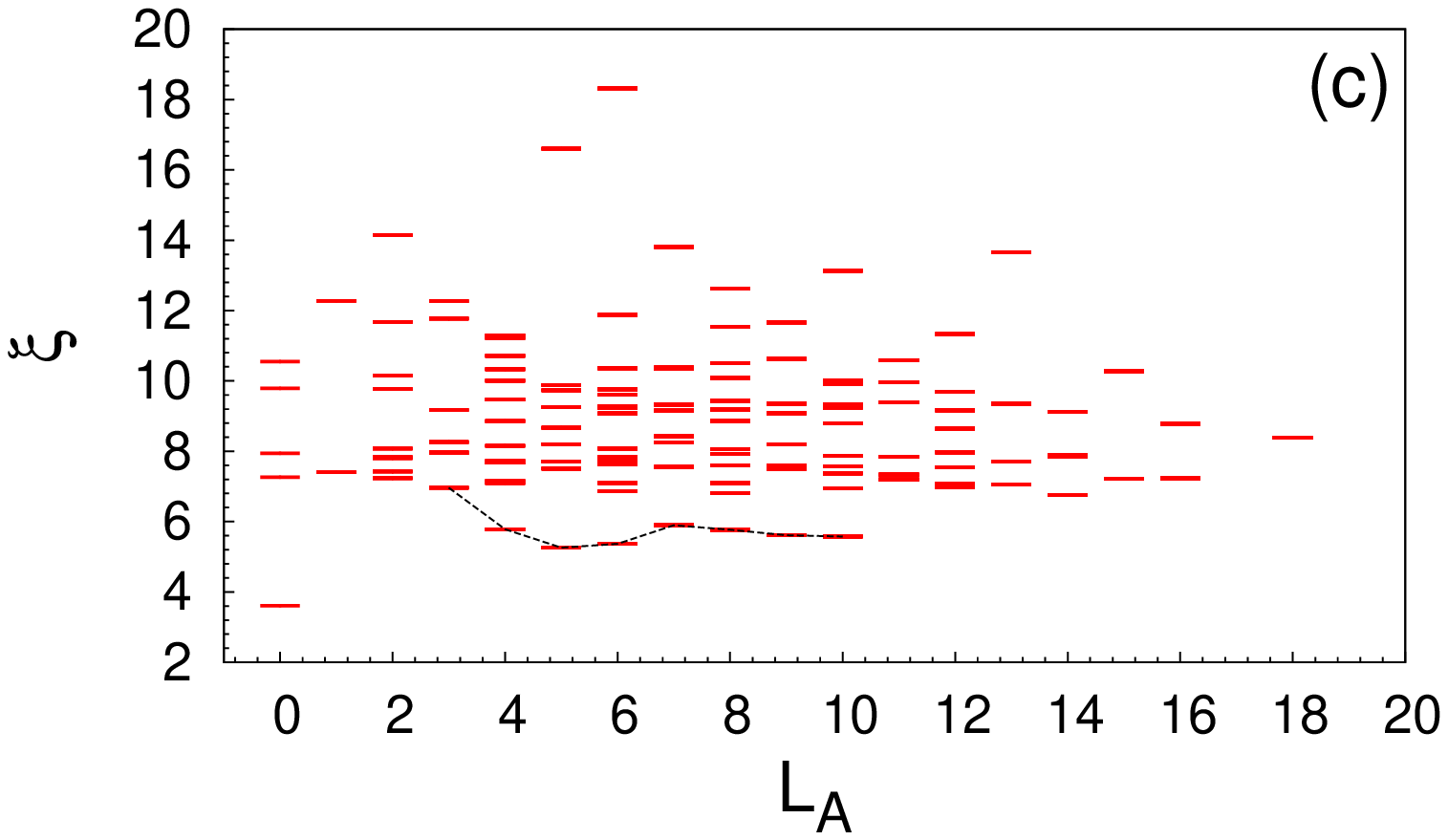}
  \end{minipage}
\caption{From left to right: (a) $\nu=1/3$ Laughlin state PES  for $N=8$ and $N_A=4$. A small size has been selected for pedagogical purposes. The counting for the $L$ multiplet is $(3,0,4,2,6,3,7,4,7,5,7,4,7,4,5,3,4,2,3,1,2,1,1,0,1)$. It matches the one expected for the $N_A=4$ Laughlin state with 12 added qhs \cite{Ardon-02jpa447,*Read-06prb245334}.(b) Coulomb ground state at $\nu=1/3$ for the same system sizes as (a). The low energy part below the dotted line is similar to the Laughlin state: the multiplet structure is identical except for one missing high energy state. (c) the bosonic MR state $\nu=1$ for $N=12$ and $N_A=6$. The dotted line is the magnetoroton-like mode, it stops at $L_A=N-2$ while the usual Laughlin magnetoroton mode ends at $L=N$.}
\label{laughlinpartessphere}
\vspace{-0pt}
\end{figure*}

We start with two well known model wavefunctions, the $\nu=1/3$ Laughlin\cite{laughlin83prl1395} (Fig. \ref{laughlinpartessphere}a) and $\nu=1$ Moore-Read (MR) states\cite{Moore-91npb362} (Fig. \ref{laughlinpartessphere}c). We chose a fermionic and a bosonic state to show that the PES properties are independent of the particle statistics. While all the PES in this paper are done for a half cut  $N_A=N/2$, all results are valid for any other cut $N_A \le E(N/2)$ where $E(x)$ is the integer part of $x$. In the general case, the counting is that of min($N_A$,$N_B$) particles with $N_\Phi$ flux quanta. We compare the counting of the $L_A$ multiplet in PES with the one expected for the Laughlin and MR states with $N_A$ particles and a total number of flux quanta $N_\Phi$ and  observe a perfect match. Moreover,  the space spread by the eigenstates of $\rho_A$ in a fixed ($L_{z_A}$, $L_A$) sector coincides with the one of the qh states in the same sector. We checked this property up to $N=13$ for the Laughlin state and $N=16$ for the MR state. These examples clearly show that the qh excitations are embedded in the ground state, and can be extracted through the PES as the  eigenstates of $\rho_A$.

We compute the PES for non-model wavefunctions such as a Coulomb ground state. Fig. \ref{laughlinpartessphere}b shows the example of $\nu=1/3$.  The low entanglement-energy part of the spectrum has the same pattern as the Laughlin state (for all $L$ sectors unlike the GES on the sphere).  While the PES does not exhibit a full entanglement gap, the Laughlin levels are separated from the spurious Coulomb ones by a visible entanglement gap for the first several levels from $L_A=24$ to $L_A=17$. The conformal limit used on the sphere to define a clear entanglement gap\cite{thomale-10pr180502} for the GES, cannot be applied for PES: such a limit breaks the rotational symmetry and spoils the multiplet structure.

In certain situations, the PES (in Fig. \ref{laughlinpartessphere}c) resembles the typical energy spectrum of the true Hamiltonian in an incompressible phase. For a half cut, in the bosonic MR state, the relation between $N_A$ and $N_\Phi$ is identical to the one of the bosonic Laughlin $\nu=1/2$ state \ie $N_\Phi=2(N_A - 1)$. The PES for the MR state features a ``ground state'' at $L_A=0$ and  a dispersing magneto-roton-like mode. The square overlap of the PES ``ground state'' and the Laughlin state is very high (from 0.9989 for $N=10$ to 0.9986 for $N=16$). Since large overlap might be misleading/accidental, we also checked that the PES and GES of the Moore-Read PES ``ground state'' resembles the one of the Laughlin state. While the Laughlin state can be thought of as a MR qh state, it is surprising that it appears incredibly close to ground state of the effective Hamiltonian related to $\rho_A$.

The Laughlin and Moore-Read states are single Jack polynomials uniquely defined by clustering conditions, whose excitations obey a unique generalized Pauli principle. One may question if the PES results are derived from these rather special properties. To settle this issue, we have considered the Haffnian state\cite{green-10thesis}. It has a known Hamiltonian for which it is the unique densest zero energy ground state but it is not uniquely defined by clustering conditions. Nevertheless, we checked up to $N=12$ particles that the Haffnian PES directly reveals all the Haffnian quasihole excitations.

It is easy to analytically show that if a wavefunction is the zero energy state of a local repulsive Hamiltonian ${\cal H}$, then eigenstates of reduced density matrix with nonzero eigenvalue (non-infinite entanglement energy) are zero modes of the same repulsive Hamiltonian but for the $N_A$ number of particles within the initial $N_\Phi$ number of fluxes. We assume that the local Hamiltonian contains at most $k$-body interactions and decompose it in ${\cal H}={\cal H}_A+{\cal H}_B+{\cal H}_{A+B}$ where ${\cal H}_A$ (resp. ${\cal H}_B$) acts only on the $A$ (resp. $B$) group of particles and ${\cal H}_{A+B}$ describes the interaction between the two groups. If $N_A<k$ then the $\ket{\Psi^A_i}$'s are trivially zero energy states of ${\cal H}_A$. If $N_A \ge k$, since ${\cal H}\ket{\Psi}=0$ then we also have ${\cal H}_A\ket{\Psi}=0$. Using the Schmidt decomposition, we deduce ${\cal H}_A\ket{\Psi^A_i}=0$. As a consequence, the number of non-zero eigenvalues of $\rho_A$ is bounded by the number of the zero energy states of the Hamiltonian for $N_A$ particles and $N_\Phi$ flux quanta. We stress that this proof does not provide an explanation for our finding that this bound is saturated for $N_A\le N/2$ for all the case studied.

The most salient feature of the PES rests in its ability to probe the qh physics even for cases where there is no known Hamiltonian. A simple example is any of the more generic $(k,r >3)$ clustered states, which, unlike the Moore-Read and Laughlin states, are not uniquely defined by their clustering properties. Since some of them are (bosonic or fermionic) Jack polynomials, they can be decomposed on the n-body basis using the recursion formula of \cite{bernevig-09prl206801}, the PES can be computed, and the results can be compared with the expected counting\cite{Ardon-02jpa447,*Read-06prb245334}. We checked that the counting matches. Moreover, we also checked that the PES eigenstate space is spanned by the $(k,r)$ Jack polynomials. All the test we have done show a perfect match between the two approaches, which suggests that the $(k,r)$ Jack polynomials do have a local Hamiltonian for which they are the densest zero energy states.

Other famous examples of wavefunctions for which no Hamiltonian is known are Jain's composite fermion (CF) hierarchical states\cite{jain89prl199}. They provide appealing explanations of many FQH features, including the series of experimentally observed fractions $p/(2p+1)$ and the compressible state at $\nu=1/2$. The CF approach maps the original problem of interacting electrons into free composite fermions (CF) - particles bound to flux quanta:

\begin{equation}
\Psi_{CF}={\cal P}_{{\rm LLL}} \prod_{i<j}\left(z_i - z_j\right)^n \Phi^{CF}_p
\label{jaincfwf}
\end{equation}

\begin{figure}[t]
\includegraphics[width=\linewidth]{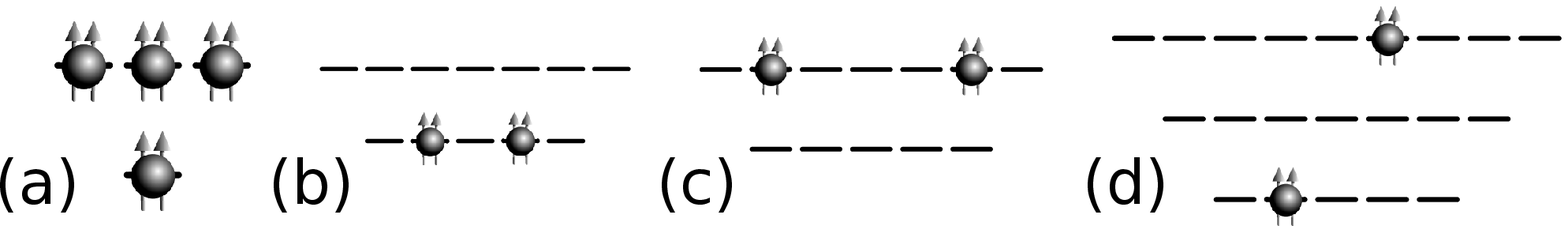}
\includegraphics[width=0.8\linewidth]{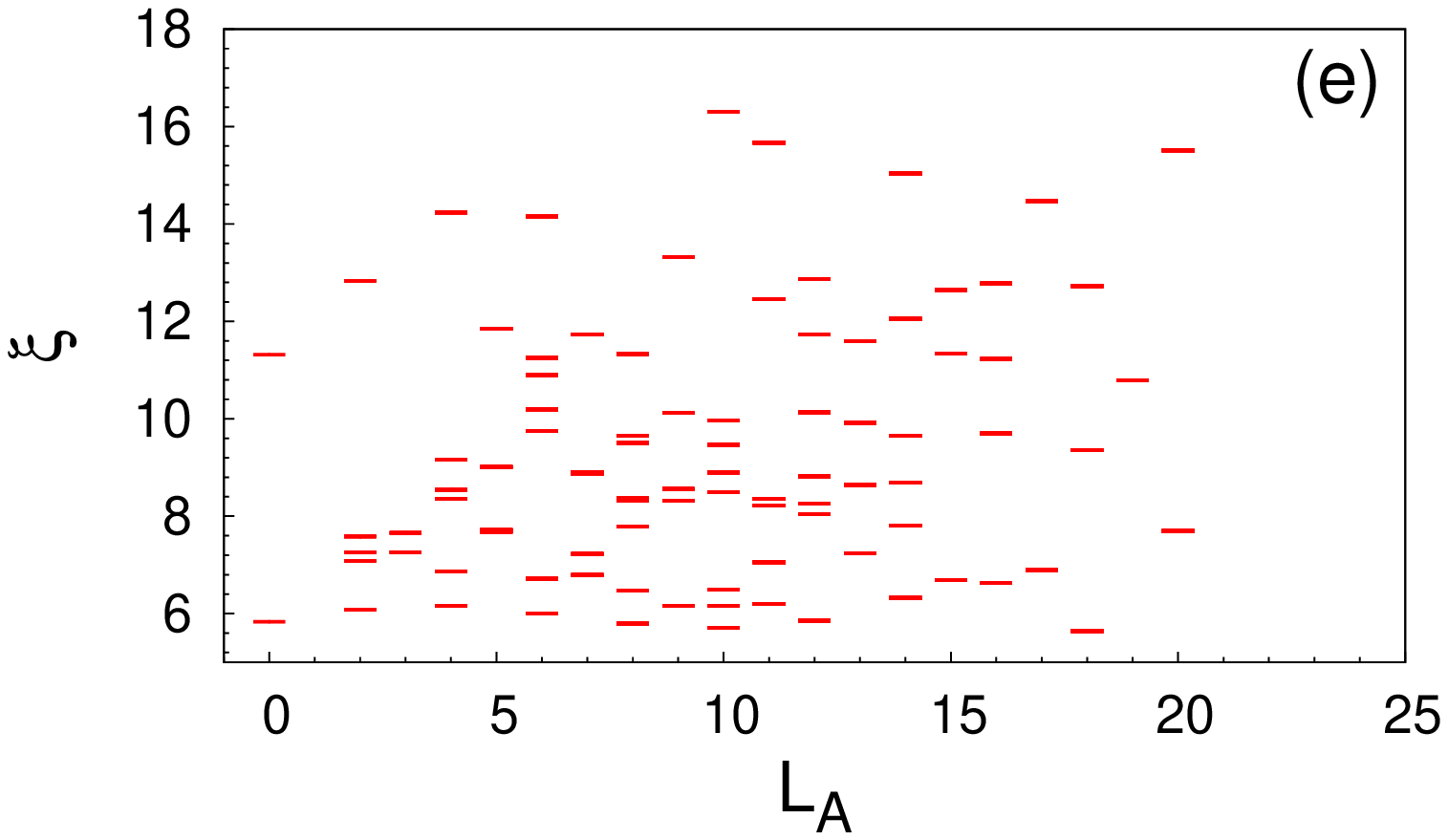}
\caption{{\it Upper part}: schematic description of the  $\nu=2/5$ CF state. Fig. (a) The ground state for 4 particles. In the PES, one removes 2 particles (2 CFs) - the 2 remaining particles feel 4 additional flux quanta. Fig. (b) The lowest effective energy configurations in this situation involve only the LLL. Figs. (c) and (d) Two other configurations having the same effective kinetic energy (not part of the lowest effective energy). Except for accidental degeneracies when the projected three $\Lambda$ levels states are  already present in two $\Lambda$ levels states, only the cases involving two $\Lambda$ levels such as shown in (c) are present in the PES. {\it Lower part}: PES for the fermionic Jain's state $\nu=2/5$ for $N=8$ and $N_A=4$. The counting matches the one expected for the $N_A=4$ CFs with $N_{\Phi,A}^*=10$ flux quanta within two $\Lambda$ levels.}
\label{jaincfpic}
\end{figure}

\noindent $\left(z_i - z_j\right)^n$  attaches $n$ flux quanta to the original particle (the $z_i$ are the $i$'th particle's complex coordinates). ${\cal P}_{{\rm LLL}}$ is the projection onto the LLL. $\Phi^{CF}_p$ corresponds to the wavefunction of the free CF in $p$ effective Landau levels, called $\Lambda$ levels ($\Lambda$L). Jain's wavefunction describing the ground state at $\nu=p/(2p+1)$ corresponds to $n=2$ and $p$ $\Lambda$L filled with CF. Before projection, for each CF configuration, we associate an effective kinetic energy assuming the $\Lambda$L are separated by an effective cyclotron energy. Physical reasoning suggests  minimizing this energy to obtain the excitations above a Jain ground state. We compare this heuristic view with the exact results based on the PES of the $\nu=p/(2p+1)$ Jain's ground states, as well as with the LLL projected CF excitation wavefunctions which we have built analytically. We have rigorously implemented Eq. \pref{jaincfwf}, applying the projection as the last step as opposed to the standard ad-hoc method used in Monte Carlo calculations \cite{jain-97ijmpb156404}. We first focus on the $\nu=2/5$ CF wavefunction involving two $\Lambda$L. We evaluate the PES up to $N=10$ (Fig. \ref{jaincfpic}e). When computing PES, keeping only $N_A$ particles amounts to keeping $N_A$ CFs. However, the flux attached to the removed particles is now felt by the remaining CFs and the number of effective flux quanta $N_\Phi^*=N_\Phi - 2 N$ changes to  $N_{\Phi,A}^*=N_\Phi^* + 2 (N - N_A)$. In the naive energy-minimizing picture, qh states above the $\nu=2/5$ ground state are obtained by considering all possible CF configurations with \emph{the lowest} total effective kinetic energy (Fig. \ref{jaincfpic}b). The counting of excitations thus obtained is lower than the one extracted from the PES. The mismatch with the PES strongly suggests this physically intuitive reasoning is wrong. 

This heuristic method neglects excitations obtained by shifting CF's from one $\Lambda$L to a higher energy $\Lambda$L. The correct way to match the PES counting and eigenstates is to consider all possible configurations in the lowest two $\Lambda$L, irrespective to their effective cyclotron energy, see Figs.\ref{jaincfpic}b,\ref{jaincfpic}c. The PES shows that the two $\Lambda$L structure is deeply encoded not only in the CF wavefunction but also in its quasihole excitations structure. The degeneracy counting of qh states is an important ingredient of the understanding of statistical properties of model states such as Laughlin or MR. PES provides an easy and analytically sound way to do the same for the CF state.

The effective cyclotron energy $\Lambda$L is crucial in reproducing the low energy structure of the Coulomb interaction, a major achievement of the CF approach, but is irrelevant for the qh structure of a given CF state. We note that a similar situation occurs in the unprojected $\nu=2/5$ CF wavefunction: this is known to be the exact densest solution of the hollow core interaction \cite{PhysRevB.44.8395} in an Hilbert space restricted to two Landau levels with a cyclotron energy set to zero. The quasihole excitations of such an interaction will also have zero energy and involve only two $\Lambda$L. However, the LLL projection is crucial in obtaining the correct counting for the excitations, as it introduces a large number of linear dependencies between the formerly unprojected quasihole states and thus reduces their counting dramatically to the correct one.

We checked that the $\nu=3/7$ Jain's state for $N=9$ has similar properties: its excitations are obtained by only involving the lowest three $\Lambda$L, and the PES gives the same counting and eigenstates as the projected CF wavefunctions. We conjecture this feature will generalize to any states of the series $p/(2p+1)$. We also analyzed the bosonic Jain wavefunctions, using single flux attachment in Eq. \pref{jaincfwf} i.e. $n=1$. We  reached larger system sizes ($N=14$ for $\nu=2/3$ and $N=15$ for $\nu=3/4$). The counting matches that of the projected CF states involving $N_A$ CFs and $N_{\Phi,A}^*=N_\Phi^* + (N - N_A)$ effective flux quanta and so for the eigenstates of $\rho_A$.

\begin{figure}[t]
\includegraphics[width=0.8\linewidth]{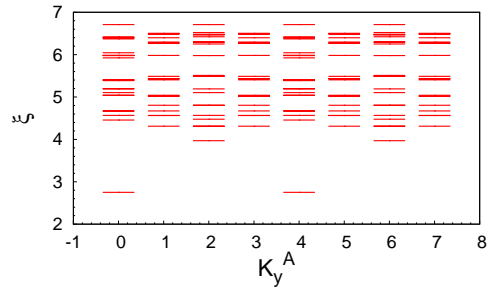}
\caption{PES of the MR state for $N=8$ particles with $N_A=4$ on the torus geometry. Due to the center of mass degeneracy, the spectrum repeats after the $K_y^A=3$ sector. As pointed out for the sphere geometry, the PES ``ground state'' is clearly separated from the excitations and is close to the Laughlin state both in terms of overlap and its own PES.}
\label{paffiantorus}
\end{figure}

Similar features of the PES hold true when considering the torus geometry. For the case of the Laughlin state, the torus GES has been recently studied in \cite{lauchli-10prl156404}. For non-abelian states, one difficulty of the torus geometry is the ground state degeneracy, which can be either of non-abelian or center of motion origin. In this situation, the definition of the total density matrix is ambiguous. In \cite{lauchli-10prl156404}, the GES was computed per momentum sector, an approach valid for the Laughlin case studied, but difficult to extend to non-abelian cases such as Moore-Read. We find that for the PES, the correct definition of the density matrix is the incoherent one summing up all sectors:

\begin{equation}
\rho=\frac{1}{d}\sum_{i=0}^{d} \ket{\Psi_i}\bra{\Psi_i}
\label{densitymatrixtorus}
\end{equation}
\noindent
where $\ket{\Psi_i}$ with $i=1,...,d$ forms an orthogonal basis of the degenerate ground state manifold ($d$ being the total degeneracy). As defined, $\rho$  commutes with the magnetic translation operators and does not depend on particular basis choice. We performed calculations using the translation symmetry along one direction. Thus our states are only labeled by the $K_y$ momentum. We checked that the PES for the bosonic Laughlin state (for N=4 to N=10) and the MR state (for N=4 to N=14) also unravels the qh physics on the torus, both in counting and in eigenstates, similar to what is obtained on the sphere geometry. Fig. \ref{paffiantorus} displays an example of the PES counting for the MR state. At half-cut, the PES has the same feature that we have mentioned for Fig. \ref{laughlinpartessphere}: a doublet ``ground state'' (due to the center of mass degeneracy) that is close to the Laughlin state and clearly separated from higher energy states.

In conclusion, we have shown that the PES allows to extract bulk excitations from a topological ground state. For the FQH effect, the PES properties are valid for model states, even in absence of an exact Hamiltonian. Future works will apply the PES to other topological phases to prove its generality.

\emph{Acknowledgements}
We thank Z. Papic, F.D.M. Haldane and especially M. Hermanns and A. Chandran for discussions.
BAB was supported by Princeton Startup Funds and the Alfred P. Sloan Foundation. BAB thanks the Ecole Normale Superieure, Paris as well as the Institute of Physics Center for International Collaboration in Beijing, China and Microsoft Station Q for generous hosting. BAB was supported by Princeton Startup Funds, Alfred P. Sloan Foundation, NSF CAREER DMR- 095242, and NSF China 11050110420, and MRSEC grant at Princeton University, NSF DMR-0819860. 

\bibliography{partes_V_2.bib}

\end{document}